\documentclass[10pt,aps,prl,twocolumn,superscriptaddress,amsmath]{revtex4-1}
\usepackage{graphicx}
\usepackage{amsmath}
\usepackage{color}

\providecommand{\rmi}{\mathrm{i}}
\providecommand{\rme}{\mathrm{e}}
\providecommand{\rmd}{\,\mathrm{d}}
\providecommand{\rmg}{\,\mathrm{g}}
\providecommand{\rmmax}{\mathrm{max}}
\providecommand{\abs}[1]{\left| #1 \right|}

\providecommand{\fref}[1]{FIG.~\ref{#1}}

\begin{document}
\title{The exact form of the Bohm criterion for a collisional plasma}
\author{Tsanko V. Tsankov}
\email[E-mail: ]{Tsanko.Tsankov@rub.de}
\author{Uwe Czarnetzki}
\affiliation{Institute for Plasma and Atomic Physics, Ruhr University Bochum, 44780 Bochum, Germany}

\date{\today}
\begin{abstract}
A long-standing debate in the literature about the kinetic form of the Bohm criterion is resolved for plasmas with single positive ion species when transport is dominated by charge exchange collisions. The solution of the Boltzmann equation for the ions gives the exact form free of any divergence and contains an additional term that is not included in the classical result. This term includes collisional and geometric effects and leads to a noticeable correction. Further, the question is addressed whether the space charge argument at the bottom of the Bohm criterion can actually lead to a meaningful definition of the transition point between bulk and sheath. The analysis is supported by a numerical model and experiments, showing excellent agreement throughout. As a novelty in diagnostics, the theoretical results allow from the ion velocity distribution function (IVDF), measured at the wall, a reconstruction of the IVDF and the electric field at any point in the plasma. This property is used to reconstruct non-intrusively also the ion density, flow velocity, mean energy and effective temperature and the electron density and temperature as functions of the spatial coordinate and potential. Finally, the fluid equation for ion momentum balance is verified.
\end{abstract}
\pacs{52.25.Dg, 
52.25.Fi, 
52.40.Kh, 
52.70.-m 
}
\maketitle

The work of Bohm~\cite{Bohm} introduced a new concept about the sheaths forming in plasmas near walls -- the space charge has to increase monotonously with the potential $\varphi$: $\partial n_{\rme}/\partial\varphi\geq \partial n_{\rmi}/\partial\varphi$, where $n_{\rme}$ and $n_{\rmi}$ are the electron and ion density. This will be termed weak form of the criterion. The strong form $n_{\rme}^{-1}\partial n_{\rme}/\partial\varphi\geq n_{\rmi}^{-1}\partial n_{\rmi}/\partial\varphi$ is derived from it and is also commonly used. For mono-energetic collisionless ions this entails that the ions attain the ion sound velocity $c_{\rmi}=\sqrt{k T_{\rme}/M}$ ($k$ is the Boltzmann constant, $M$ the ion mass and $T_{\rme}$ the electron temperature) before entering the sheath~\cite{Oksuz02}. Coincidentally, at this velocity the quasi-neutral solutions for the plasma bulk break down, promoting the ubiquitous use of this result as a boundary condition for numerical simulations and sheath models. The concept of the Bohm criterion has been also extended to account for multiple charged species~\cite{Riemann95} as well as for species possessing arbitrary velocity distribution functions~\cite{Boyd59,Harrison59,Allen09}. For the case of single species of positive ions and electrons the criterion is $\left<\varepsilon_{\rmi}^{-1}\right>\leq \left<\varepsilon_{\rme}^{-1}\right>$~\cite{Riemann91}. The brackets stand for averaging over the velocity distribution functions of electrons $f_{\rme}$ and ions $f_{\rmi}$ and $\varepsilon_{\rme,\rmi}$ is their kinetic energy. In the following we denote $\left<\varepsilon_{\rme}^{-1}\right>=2 B_{\rme}$. The sheath edge is associated with the point where the condition is marginally satisfied~\cite{Riemann91}. This formulation of the sheath criterion contains a divergence for the ionic part \cite{Baalrud11b} and does not give a meaningful result at elevated pressures~\cite{Brinkmann11,Valentini15}. 

The treatment here removes this divergence and adds an additional term that restores the validity of the criterion for any collisionality. Furthermore, the general concept of a Bohm criterion is discussed and the analysis is supported by a numerical simulation and experiment, involving also a new diagnostic concept.

The distribution of the electrons near the sheath edge is nearly isotropic due to the repelling sheath and $\left<\varepsilon_{\rme}^{-1}\right>$ is well defined. Owing to the accelerating field near the sheath edge the distribution of the ions is mostly one-dimensional. When ions with zero velocity are present $\left<\varepsilon_{\rmi}^{-1}\right>$ diverges. The common assumption~\cite{Riemann12,Baalrud12,Kos15} $f_{\rmi}(v=0)=0$ does not hold for charge-exchange (CX) collisions typical for e.g.\ atomic ions in their parent gas. This is visible from the solution of the Boltzmann equation (BE) including collision operators for charge-exchange collisions (with a constant mean free-path $\lambda$):
\begin{equation}
	v \frac{\partial}{\partial r} f_{\rmi} + \kappa \frac{v}{r} f_{\rmi} +  \frac{e}{M} E \frac\partial{\partial v} f_{\rmi} = -\frac{\abs{v}}{\lambda} f_{\rmi} +  \delta(v) Q(r).
	\label{eq:BE}
\end{equation}
\noindent Here $E(r)$ is the electrostatic field and $e$ is the elementary charge. The equation determines the ion velocity distribution function (IVDF) $f_{\rmi}(v, r)$, defined for a single spatial $R \geq r \geq 0$ and velocity coordinate $v\geq 0$ for simple geometries (plane: $\kappa=0$, cylinder: $\kappa=1$, sphere: $\kappa=2$). Ionization by electrons (ionization frequency: $\nu_{iz}(r)$ with cross section from \cite{lxcat}) determines the source term
\begin{equation}
	Q(r)=\int \frac{\abs{v}}{\lambda} f_{\rmi} \rmd v + \nu_{iz} n_{\rme} =  \frac{1}{\lambda r^{\kappa}} \int\limits_{0}^{r} \nu_{iz}n_{\rme} \tilde{r}^{\kappa} \rmd \tilde{r} + \nu_{iz} n_{\rme}.
\end{equation}
\noindent The solution is obtained with the \textit{Ansatz} $f_{\rmi}(v,r)=g(v,r)\Theta(v)\Theta(v_\rmmax(r)-v)$:
\begin{equation}
	g_{\rmi}(v(r,r^{\prime}),r) = \frac{M}{e} \frac{Q(r^{\prime})}{E(r^{\prime})} \left(\frac{r^{\prime}}{r}\right)^\kappa  \exp\left(-\frac{r-r^{\prime}}\lambda\right).
	\label{eq:BEsol}
\end{equation}
\noindent Ions created at a position $r^{\prime}\geq 0$ reach a position $r \geq r^{\prime}$ at an energy of $\varepsilon = \frac12 M v^{2} = e\left(\varphi(r^\prime) - \varphi(r)\right)$. Therefore, $v=0 \leftrightarrow r^{\prime} = r$ and $v=v_{\rmmax}(r) \leftrightarrow r^{\prime} = 0$ and $r^{\prime}(v, r)$ is a function of the total energy. Further, clearly $g(v=0,r)= \frac{M}{e E(r)}Q(r) \neq 0$. 

\begin{figure*}
	\includegraphics[width = 0.32\textwidth]{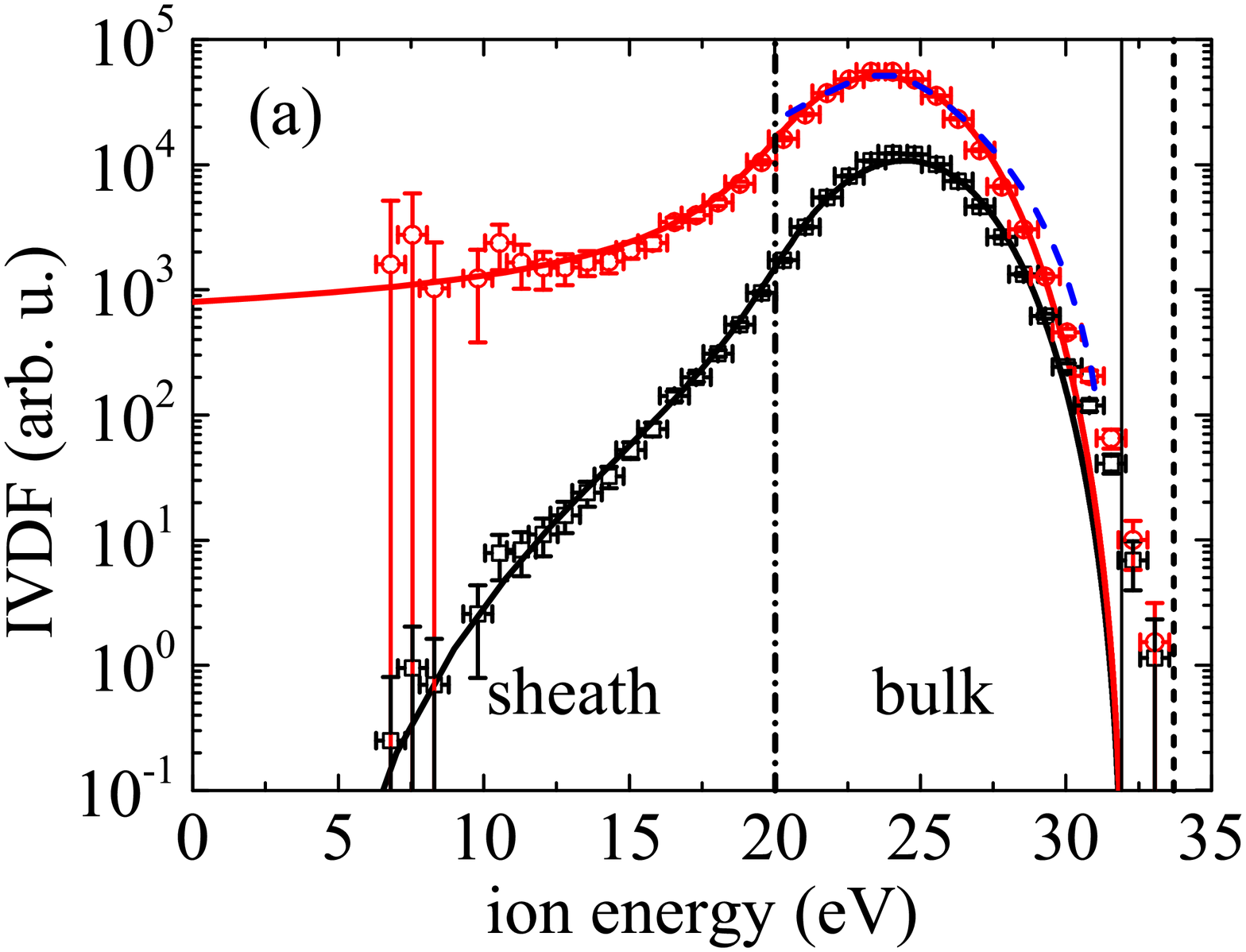}\hspace{1em}
	\includegraphics[width = 0.27\textwidth]{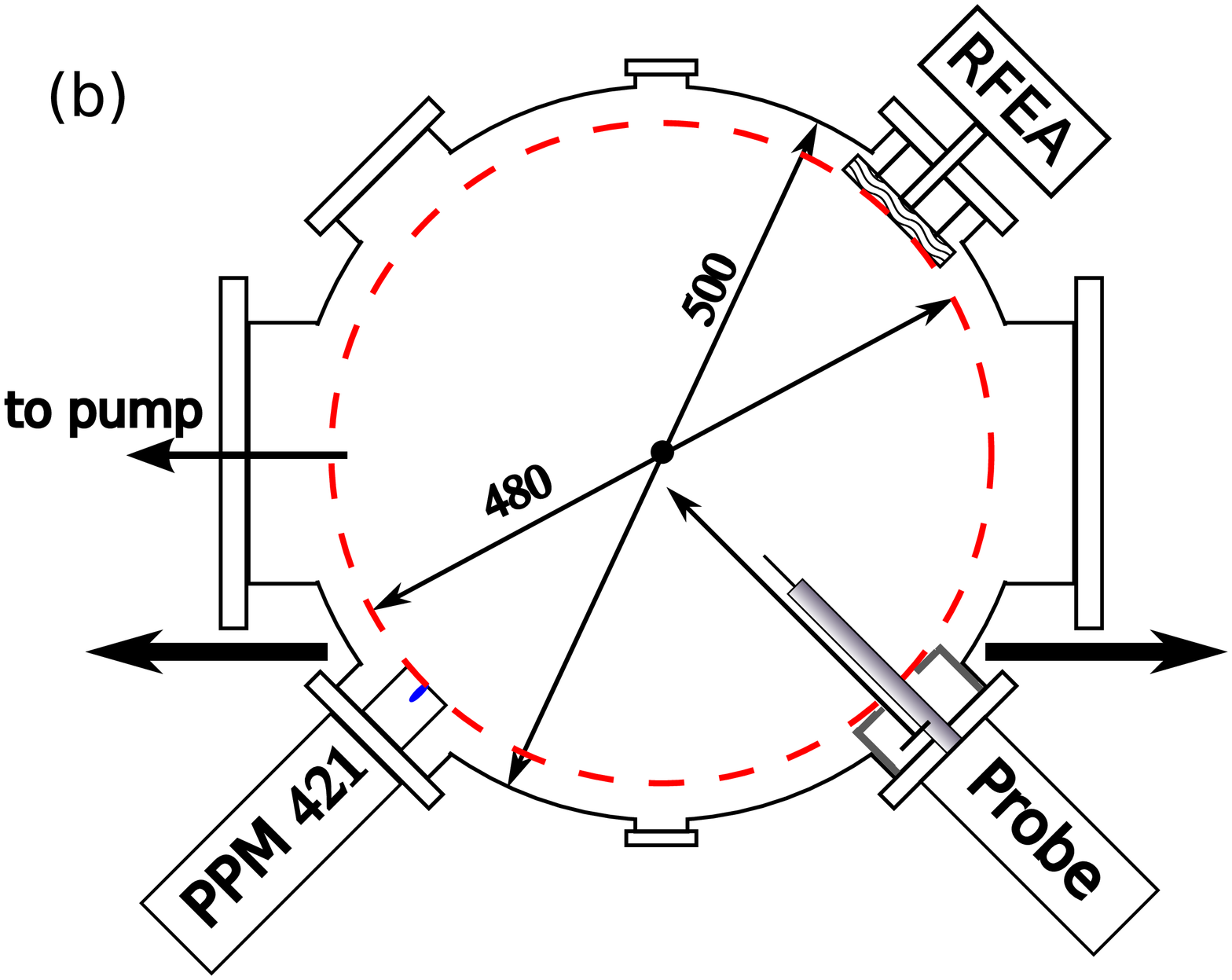}\hspace{1em}
	\includegraphics[width = 0.32\textwidth]{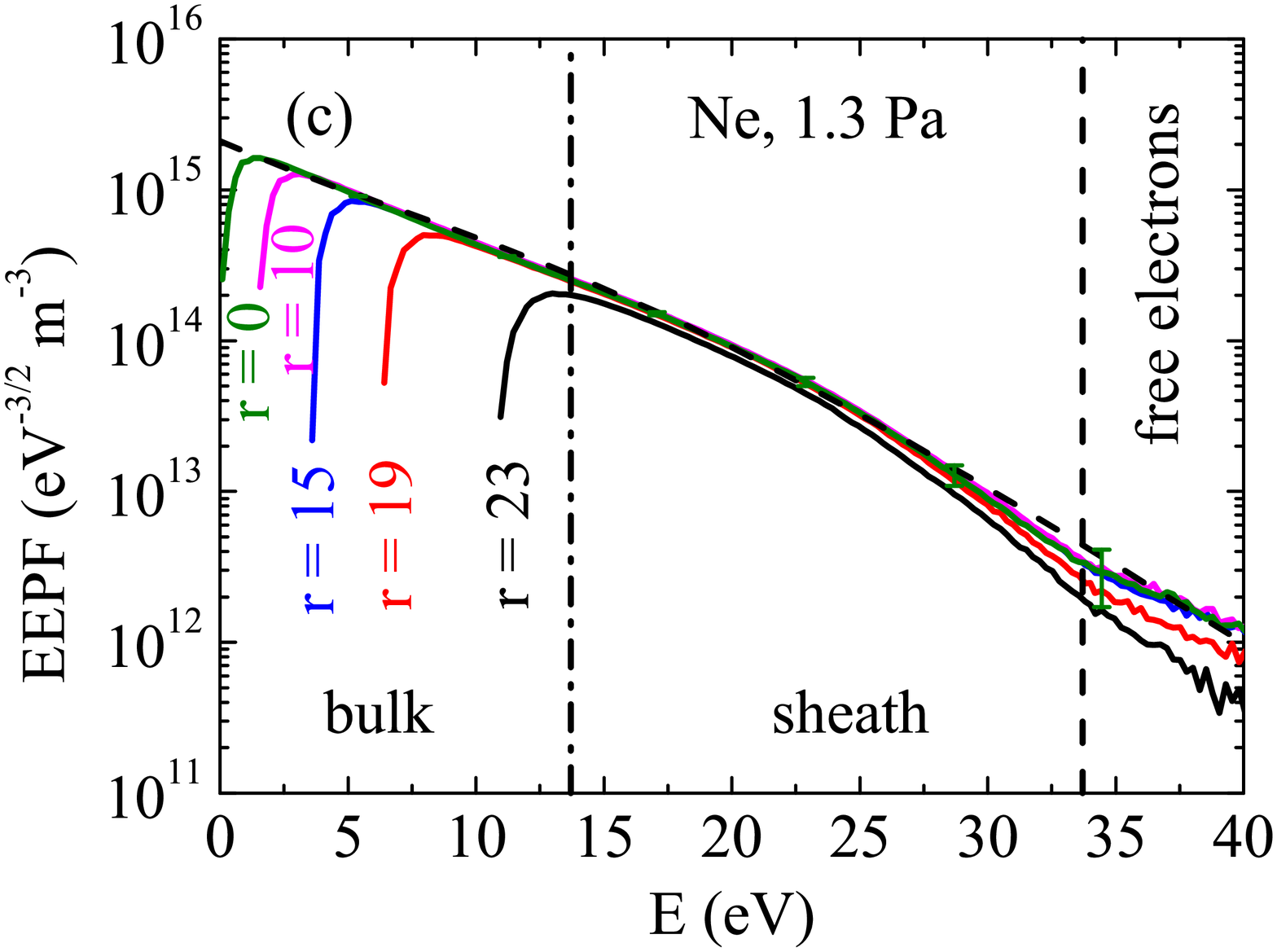}
	\caption{\label{fig:setup} (a) Measured (symbols) and calculated (curves) IVDF corrected (red) and uncorrected (black) for the depletion at low energies. IVDF from probe data (dashed blue curve) is also shown in the range 21 to 32~V. (b) Setup of the diagnostics (chamber dimensions in mm). (c) Measured EEPF at selected radial positions (in cm) and shifted by the local value of the plasma potential $V_{p}$. The dashed black curve is the parametrized $f_{\rme 0}$.  A dash-dotted line marks the energy where $u=c_{\rmi}$ (\fref{fig:Results}(c)). The energy is referenced to the grounded chamber walls for the ions in (a) and to $V_{p}(r=0)$ for the electrons in (c). The potential drop between plasma axis and walls is indicated by a dashed line (experimental value) and a continuous line (value from the model). The difference of about 1~V is within the margin of the uncertainty due to assumptions in the model and experimental errors.}
\end{figure*}

From \eqref{eq:BE} the strong form of the Bohm criterion becomes $B_{\lambda} + B_{\rmi} \leq B_{\rme}$ where a very small term containing $g(v_\rmmax)/(M v_{\rmmax})$ has been neglected. The two ion terms are:
\begin{eqnarray}
	B_{\lambda} = \frac{1 + \kappa\lambda/r}{e E \lambda} = \frac{g(v=0,r)}{M Q(r)\lambda}\left(1+\kappa\frac{\lambda}{r}\right) \nonumber \\*
	\approx \frac{g(v=0,r)}{\int_{0}^{v_\rmmax} M v\, g \rmd v}\left(1+\kappa\frac{\lambda}{R}\right), 
	\label{eq:Blambda}
\end{eqnarray}
\begin{equation}
	B_{\rmi} = \left. \int\limits_{0}^{v_\rmmax}\frac{1}{M v} \frac{\partial g}{\partial v} \rmd v \middle/ \int\limits_{0}^{v_\rmmax}g \rmd v \right..
	\label{eq:Bi}
\end{equation}

\noindent The term $B_{\rmi}$ corresponds to the classical ion term found in the literature but without the divergence, that results only from an invalid partial integration. The new term $B_{\lambda}$ is of key importance especially under conditions of high collisionality or strong geometrical effects, e.g.\ around probes. The importance  of this term under collisional conditions is demonstrated here by experiments and a model. 

The model with cylindrical symmetry ($\kappa=1$) aims at simulating the experiment described subsequently. 
The model solves numerically the continuity and momentum equation for ions and Poisson's equation along the radial coordinate $r$ for the ion flow velocity $u$, the ion density $n_{\rmi}$, and the electric field $E$. Further, the electric field is integrated to yield the plasma potential $\varphi$. From these data then the IVDF is calculated via equation \eqref{eq:BEsol} and used for determination of the kinetic moments. The electron density $n_{\rme}$, temperature $T_{\rme}$, and ionization rate $v_{iz}$ are calculated from the measured non-local EEPF (\fref{fig:setup}) as a function of the plasma potential and serve as an input to the ion equations. Note that the only input to the calculated IVDF from the fluid model is the electric field / potential. The ion momentum equation is:
\begin{equation}
	u \partial u/\partial r = e E/M - \pi/(2\lambda) u\sqrt{u^{2} + u_{z}^{2}} -\nu_{im}u - \nu_{iz} u n_{\rme}/n_{\rmi}.
	\label{eq:ui}
\end{equation}
\noindent The factor $\pi/2$ in the CX momentum loss term has been discussed in the literature \cite{Czarnetzki13} and is further discussed here in comparison with the experiment. In addition to charge exchange ($\lambda = 1.52$~cm~\cite{Jovanovic02}) also elastic collisions ($v_{mi}= 5.4\ 10^{4}$~s$^{-1}$~\cite{Jovanovic02}) and a finite axial velocity ($v_{z} = 640$ m/s, 12\% of $c_{\rmi}$ on the axis, $c_{\rmi 0}$) equivalent to the ion thermal velocity are taken into account. These effects are significant only at very low energies in the central part of the plasma ($r<R/2$). In the IVDF at the wall this corresponds to ions with $\varepsilon_{\rmi} > 30$ eV (\fref{fig:setup}). Further, in the radial continuity equation the axial flow is taken into account by a reduced ionization frequency $\nu_{izr} = \nu_{iz} (1 - \eta)$. The parameter $\eta$ is adjusted so that flux balance is reached at the wall at $r = R$. Indeed a reasonable value of $\eta = 0.455$ is found. The neutral gas density determining the values of all collision parameters follows from the measured neon pressure $p = 1.3$ Pa and the gas temperature $T_{\rmg} = 400$~K. The latter value is estimated from earlier measurements in argon~\cite{Celik11} accounting for the different heat conductivities of the gases. Initial conditions on the axis ($r=0$) are zero ion velocity $u$, electric field and potential $\varphi$. The ion density in the center, $n_{0}=3.0\ 10^{16}$~m$^{-3}$, is from the experiment. 

For the experiments an inductively coupled plasma in neon at $p=1.3$ Pa and a power of 600~W is used. Details about the cylindrical discharge chamber ($R=25$~cm) can be found in~\cite{Celik12PREa}. The measurements are in a horizontal plane 24~cm below the inductive antenna (\fref{fig:setup}(b)): plasma bulk parameters by a home-made Langmuir probe, LP, (tip length and radius 9 and 0.05~mm) and mass-selected IVDF by a Balzers Plasma Process Monitor, PPM. The calibration of the energy axis is checked against a Retarding Field Energy Analyser, RFEA. The entrance orifices of both devices are 1~cm closer to the chamber axis. The probe enters the chamber through an orifice of 2~cm in diameter.

The measured IVDF for a mass 20 amu (\fref{fig:setup}(a)) shows depletion of low energy ions. This is corrected by an energy-dependent factor $\psi(\varepsilon) = \cosh^{-2}\left(\sqrt{\varepsilon_c/\varepsilon}\right)$, derived from a model that assumes a homogeneous ion beam of density $n_{\rmi}$ inside the PPM that traverses a length $L$. The adjustable parameter $\varepsilon_{c}= e^{2}n_{\rmi} L^{2}/(4 \varepsilon_{0})$, with $\varepsilon_{0}$ the permittivity of vacuum, gives for $\varepsilon_{c}=408$~eV excellent agreement with the IVDF reconstructed from the fluid model (\fref{fig:setup}(a)). Deviations between the measured and the simulated IVDF can be noticed for $\varepsilon_{\rmi}>30$ eV which is most likely related to elastic collisions and a finite axial velocity as discussed above.

Further on, moments are shown only for $\varepsilon_{\rmi}> 7$ eV, since IVDF data points are missing at lower energies. For the evaluation of quantities from the IVDF the translation property of the solution \eqref{eq:BEsol} is used: $g(\varepsilon, r(\varphi)) = g(\varepsilon-e\varphi, R) \left(r/R\right)^{\kappa}\exp\left[(R-r)/\lambda\right]$. The obtained quantities are a function of the potential shift $\varphi$. The electric field is obtained from \eqref{eq:Blambda}. In the range of validity of both diagnostics (21 to 32 eV) the electric field measured by the LP via the plasma potential deviates from the one derived from the IVDF by not more than 13\%. 
This allows also construction of part of the IVDF at the wall from the LP data in the bulk with excellent agreement (\fref{fig:setup}(a)).
 The spatial coordinate as a function of the potential results from $x = -\int \rmd \varphi/E$. Since the integrand requires also data starting from zero energy, extrapolation is used between $\varepsilon = 0$ and $\varepsilon = 7$~eV. However, this range contributes only marginally. The measurement of only relative ion fluxes requires calibration of the ion density, made via the LP data at $V_p = 22$~V ($r = 23.5$~cm). 

The electron energy probability functions, EEPF~\cite{Godyak92}, obtained with the LP by the Druyvesteyn method show non-local behaviour~\cite{Kortshagen94} $f_{\rme} = f_{\rme0}(\varepsilon_{\rme} - e\varphi)$ (\fref{fig:setup}(c)), expected at low pressure (electron energy relaxation length much larger than $R$). The envelope of all distribution functions, $f_{\rme0}$, is parametrized and used to calculate the electron quantities as function of the potential.

The uncertainties in the experimental data are estimated from the uncertainties in the measured IVDF and the precision of the calibration. Typically they are in the range 5 to 10\% and often the error bars do not exceed the size of the symbols in the graphs. The precision of the LP data are estimated from the resolution of the probe system (16 bit in current and voltage) and are below 5\%.

\begin{figure}[t!]
	\includegraphics[width=0.9\columnwidth]{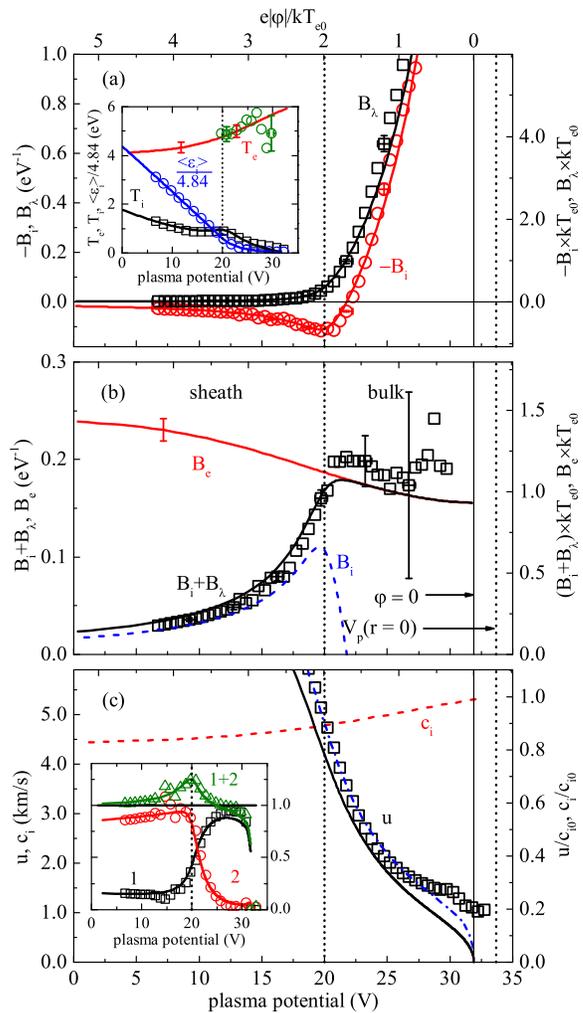}
	\caption{\label{fig:Results} (a) The ion terms $B_{\rmi}$ (red) and $B_{\lambda}$ (black) in the strong form of the Bohm criterion obtained from the experimental (symbols) and calculated (lines) IVDF. Inset: electron temperature from LP (red curve) and from experimental IVDF (open green circles), effective ion temperature $T_{\rmi}$ (black) and mean ion energy (blue) $2 \left<\varepsilon_{\rmi}\right> / [\ln\left(M/(2\pi m_{\rme})\right) +1]$ from model (curves) and experiments (symbols). (b) Ion, $B_{\rmi}$ and $B_{\rmi}+B_{\lambda}$, and electron component, $B_{\rme}$. (c) Ion flow velocity from measured (symbols) and calculated (dash-dotted blue curve) IVDF  together with fluid velocity (black line) and ion sound speed (dashed red curve) from LP. Inset: model (curves) and experimental (symbols) values of the terms in the ion momentum balance (1 -- $\pi M u^{2}/(2 e E \lambda)$; 2 -- $\partial (M u^{2}/2)/(\partial e\varphi)$). The bottom and left axes show absolute values and the top and right axes are in normalized units. The vertical lines are as in \fref{fig:setup}.}
\end{figure}

The terms in the strong Bohm criterion, obtained both from the experimental IVDF and EEPF and from the reconstructed IVDF from the fluid model, are compared in \fref{fig:Results}. The new ion term $B_{\lambda}$ is comparable in magnitude to the classical term $B_{\rmi}$, revealing the importance of its addition. This becomes clear also in \fref{fig:Results}(b) where $B_{\rmi}$ and $B_{\rme}$ do not have a crossing point and a definition of a sheath edge is not possible. With the inclusion of $B_{\lambda}$ the ionic component becomes nearly equal to the electronic one at about the position where $u=c_{\rmi}$ (marked by a vertical line). It is also remarkable that $B_{\rmi}$ has a maximum at nearly the same position. In the plasma bulk ($u<c_{\rmi}$) the two ionic terms have opposite signs and are large in magnitude. Therefore, their sum is experimentally very sensitive to noise, hence the large error bars. There is also a slight difference in the value of the potential over the plasma obtained experimentally and from the model. The difference is about 5\% and stems from the uncertainty in the ion friction and in the electron temperature.   

From the experimental ion density $n_{\rmi}(\varphi)$ and electric field $E(\varphi)$ the electron density $n_{\rme}(\varphi)$ can be obtained via Poisson's equation. This allows an estimate for the electron temperature to be obtained via $k T_{\rme}/e = n_{\rme} (\partial n_{\rme} / \partial \varphi)^{-1}$, which is based on assuming Boltzmann distributed electrons. These values show remarkable agreement with the values from LP data (inset of \fref{fig:Results}(a)). Naturally, this delicate analysis is limited to the region around the maximum of the IVDF.

The IVDF delivers also information on the effective ion temperature $k T_{\rmi} = M\left(\left<v^{2}\right> - u^{2}\right)$ and the mean energy $\left<\varepsilon_{\rmi}\right> = M\left<v^{2}\right>/2$ (inset of \fref{fig:Results}(a)). The mean energy increases nearly linearly with the potential in the sheath (very weakly collisional sheath) and reaches the expected value of $ k T_{\rme}\left[\ln\sqrt{M/(2\pi m_{\rme})} + \frac12\right] \approx 4.84 k T_{\rme}$ ($m_{\rme}$ is the electron mass) at the wall. This value follows from the balance of ion and electron fluxes at the wall. The result demonstrates again the consistency of experiment and model. The effective ion temperature increases in the plasma bulk due to ion collisions.  $T_{\rmi}$ increases also in the sheath due to small but finite friction as can be seen in the analysis of its contribution to the momentum equation (inset of \fref{fig:Results}(c)).

\begin{figure}
	\includegraphics[width=.8\columnwidth]{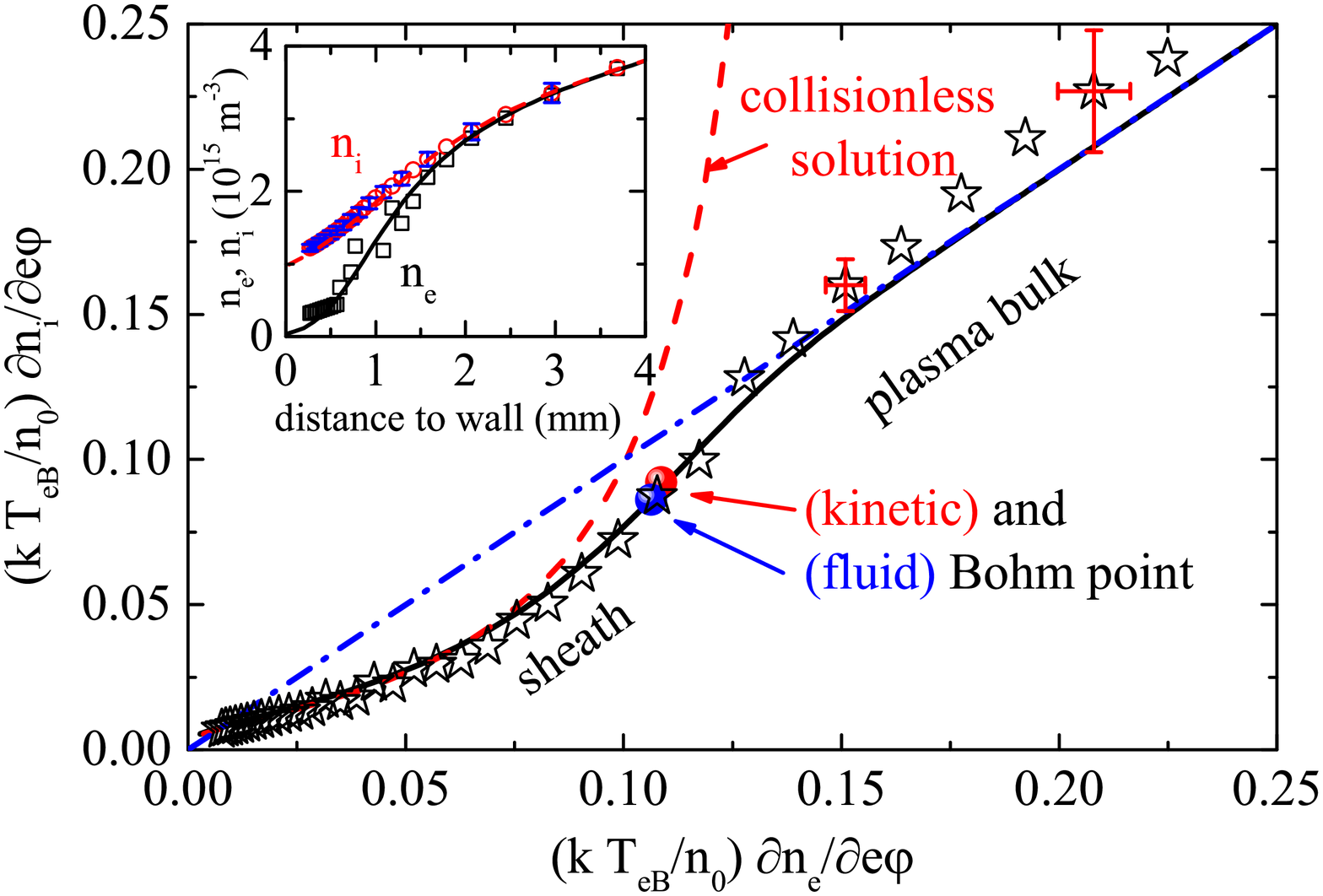}
	\caption{\label{fig:Bohm}Weak form of the Bohm criterion: the diagonal corresponds to both terms being equal (the bulk) and in the sheath the ion part drops below the value for the electron part. Shown is also the collisionless sheath expression (red dashed curve) for the ions. The symbols are the values from the measurements. Inset: ion (red) and electron (black) densities in the vicinity of the wall from model (curves) and measurements through the IVDF (symbols).}
\end{figure}

The validity of the major approximations in the fluid model can also be tested using the experimental data. The inset of \fref{fig:Results}(c) shows a comparison of the ratio of the leading terms in the ion momentum balance. Curve and symbols denoted as 1 are the ratio of the CX momentum loss term to the electric force term. Curve and symbols 2 show the contribution of ion inertia relative to the electric force. To our knowledge this is the first direct experimental test of the fluid equations that reveals the expected behaviour -- the ion friction is dominant in the plasma bulk and the ion inertia controls the sheath. The ratio of the sheath width $s\approx 0.2$ cm (\fref{fig:Bohm}) to $\lambda$ is 0.13. Accordingly, CX friction contributes about 13\% to the momentum balance. 

The sum of the two ratios is larger than 1 by about 20\% in the transition region between bulk and sheath. Here the factor $\pi/2$ used in the fluid equation \eqref{eq:ui} underestimates the CX-friction. Indeed, the factor is exact only in homogenous fields and the large gradients in the transition region lead to small but noticeable deviations. This observation agrees well with the discussion in \cite{Czarnetzki13}. As a consequence fluid and kinetic flow velocities deviate slightly in \fref{fig:Results}(c). Nevertheless, the error is small since the spatial region extends only over a few $\lambda_{D}\ll \lambda$ ($\lambda_{D}=0.3$ mm is the Debye length at the Bohm point).

Finally the suitability of the Bohm criterion for providing a condition for the sheath edge needs to be discussed. In \fref{fig:Bohm} the weak form of the Bohm criterion is presented. For convenience both axes are normalized by $k T_{\rme B} / n_{0}$, $k T_{\rme B}$ is the electron temperature at the point where $u = c_{\rmi}$. For a homogeneous electron temperature the abscissa would be identical to the relative electron density and both scales would run from zero to one. The simulation results are well confirmed by the experimental values which are constructed from the IVDF (ionic part) and the EEPF (electronic part). The figure shows the obvious fact that in a plasma always $n_{\rme} < n_{\rmi}$, even in the center. Therefore, the derivatives are identical only in the center and far into the sheath. The latter identity results from the fact that space charge has a maximum when the electron density has effectively vanished but the ion density has not decreased too much. However, this point in the lower left corner is apparently unsuitable for the purpose of defining a transition point from the quasi-neutral bulk to the sheath.

On the other hand, the quasi-neutral fluid equations for the bulk yield under any collisionality a divergence at the ion sound speed in the derivatives of the potential, the density, and the ion velocity as can be seen e.g.\ from the velocity equation $u^{\prime} \propto (c_{\rmi}^{2}-u^{2})^{-1}$. Traditionally, the ion sound speed $c_{\rmi}$ is identified as the Bohm speed $u_{B}$ by assuming collisionless ions and neglecting ionization in the sheath. The corresponding solution is also shown in \fref{fig:Bohm}:
\begin{equation}
	\frac{k T_{\rme B}}{n_{0}} \frac{\partial n_{\rmi}}{\partial e \varphi} = \frac{n_{B}/n_{0}}{\left[1-2 e\Delta \varphi/(k T_{\rme B})\right]^{3/2}}.
\end{equation}
The potential difference to the Bohm point $\Delta\varphi$ and the ion density ratio $n_{0}/n_{B} =8.65$ are taken from the model. Apparently the interception between this formula and the diagonal, which corresponds to the equality in the classical Bohm criterion, is well off the correct solution. Further, the above formula has an inherent divergence at $\Delta\varphi = k T_{\rme B}/2$ which always leads to an interception, i.e.\ to a solution. However, this solution is clearly incorrect. In reality, for any finite collisionality and ionization, the transition between the sheath and the bulk region is smooth and gradual and there is no interception anywhere. The curve comes very close to the diagonal but it never intercepts it so that the Bohm criterion never applies with an equal sign. Solutions are only found by using it with an incorrect formula.

In conclusion, the existing kinetic form of the Bohm criterion has been corrected by removing the inherent divergence in the ionic term. Consistent derivation shows that there exists a second term, that stems from collisions and geometry effects. Without it, the equality sign in the Bohm criterion, that defines the sheath edge position, can not be satisfied. Analysis shows that even then the equality holds only approximately and strictly speaking the Bohm criterion does not define the sheath edge. The condition $u=c_{\rmi}$ still remains a meaningful definition of the sheath edge, but it no longer follows from the classical Bohm criterion.

The authors are grateful to Prof.\ Satoshi Hamaguchi for several inspiring discussions over the course of this investigation. Further, the authors want to note that they have contributed equally to this work.

\bibliography{refs}

\end{document}